\renewcommand{\Re}{\mathop{\rm Re}\nolimits}
\documentstyle[amsmath,epsfig]{prhep97}

\makeatletter
\let\chapter\hid@chapter
\makeatother

\begin{document}
%\pagenumbering{empty}

% The following definitions need to be customised;

% Will appear on page headings
\authorrunning{Robin\,G.\,Stuart}
\titlerunning{{\talknumber}: \({\cal O}(N_f\alpha^2)\)
                                Radiative Corrections}

% Now the full name of author and talk

% For plenary talks, the talk number is that of the session
\def\talknumber{703}

\title{{\talknumber}: \({\cal O}(N_f\alpha^2)\)
                      Radiative Corrections in
                      Low-Energy Electroweak Processes}
\author{Robin\,G.\,Stuart
(stuartr@umich.edu)}
\institute{ Randall Laboratory of Physics,
            500 E. University, Ann Arbor, MI 48109-1120, USA}

\maketitle

\begin{abstract}
Of the the three best-measured electroweak observables, $\alpha$, $G_\mu$
and $M_Z$, the first two are extracted from low-energy processes. Both
$G_\mu$ and $M_Z$ are now known to an accuracy of about 2 parts in $10^5$
and there is a proposal to improve the measurement of the muon lifetime
by a factor of 10 in an experiment at Brookhaven. Yet calculations of
electroweak radiative corrections currently do no better than a few parts
in $10^3$--$10^4$ and therefore cannot exploit to available experimental
precision. We report on the calculation of the ${\cal O}(N_f\alpha^2)$
corrections to Thomson scattering and the muon lifetime from which
$\alpha$ and $G_\mu$, respectively, are obtained. The ${\cal O}(N_f\alpha^2)$
corrections are expected to be a dominant gauge-invariant subset of
2-loop corrections.
\end{abstract}
%
%\section{\({\cal O}(N_f\alpha^2)\) Radiative Corrections}

We have studied the ${\cal O}(N_f\alpha^2)$ corrections to Thomson
scattering and muon decay \cite{loweng,thomson}.
and present some of our results below.
These are corrections that come from 2-loop diagrams containing a
massless fermion loop where $N_f$ is the number of fermions.
Because, $N_f$ is quite large, the ${\cal O}(N_f\alpha^2)$
corrections are expected to be a dominant subset of 2-loop graphs
contributing at about the $1.5\times10^{-4}$ level.
Moreover, since $N_f$ provides a unique tag, the complete set of
${\cal O}(N_f\alpha^2)$ corrections contributing to a particular
physical process will form a gauge-invariant set.

The vast majority of ${\cal O}(N_f\alpha^2)$ diagrams, including box
diagrams, can be reduced to expressions involving a master integral
of the form
\begin{multline*}
\int\frac{d^np}{i\pi^2}\frac{1}{[p^2]^j[p^2+M^2]^k}
\int\frac{d^nq}{i\pi^2}\frac{1}{[q^2]^l[(q+p)^2]^m}\\
    =\frac{\pi^{n-4}}{(M^2)^{k+j+l+m-n}}
          \Gamma\left(l+m-\frac{\textstyle n}{\textstyle 2}\right)
          \Gamma\left(\frac{\textstyle n}{\textstyle 2}-l\right)
          \Gamma\left(\frac{\textstyle n}{\textstyle 2}-m\right)\\
    \times
    \frac{\Gamma(n-j-l-m)\Gamma(k+j+l+m-n)}
         {\Gamma\left(\frac{\textstyle n}{\textstyle 2}\right)
          \Gamma(k)\Gamma(l)\Gamma(m)\Gamma(n-l-m)}
\end{multline*}
using techniques described in \cite{loweng}.

There are a few diagrams in which the internal fermion mass cannot be set
to zero initially but they can be treated via a mass expansion along
the lines given in ref.\cite{DavyTausk}.

Dimensional regularization is employed throughout and we find that
simpler results are often obtained by carrying the full analytic dependence
on $n$, the number of space-time dimensions, rather than expanding up to
finite terms in $\epsilon=2-n/2$. A case in point is the $Z$-$\gamma$
mixing at zero momentum transfer that contributes to Thomson scattering
and for which the diagrams are shown in Fig.1.
\begin{figure}[h]
\vglue -0.7cm
\begin{center}
\epsfig{file=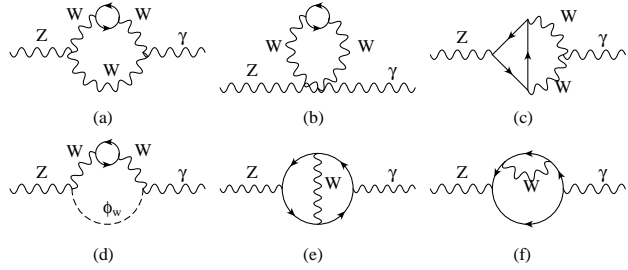,width=3.3in}
\end{center}
\vglue -0.4cm
\caption{${\cal O}(N_f\alpha^2)$ corrections to the
         $Z$-$\gamma$ mixing, $\Pi^{(2)}_{Z\gamma}(0)$.}
\vglue -0.4cm
\end{figure}

The result is
\[
\Pi_{Z\gamma}^{(2)}(0)=
 \left(\frac{g^2}{16\pi^2}\right)^2 8 s_\theta c_\theta M_Z^2
 \frac{(\pi M_W^2)^{n-4}}{n}
\Gamma(4-n)\Gamma\left(2-\frac{n}{2}\right)\Gamma\left(\frac{n}{2}\right)
\]
where $s_\theta$ and $c_\theta$ are $\sin\theta_W$ and $\cos\theta_W$
respectively and we assume one generation of massless fermions.

The diagrams contributing at ${\cal O}(N_f\alpha^2)$ to the photon
vacuum polarization are shown in Fig.2.
\begin{figure}
\vglue -0.7cm
\begin{center}
\epsfig{file=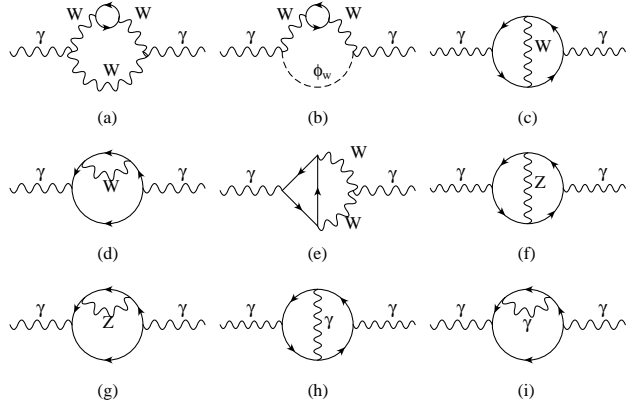,width=3.3in}
\end{center}
\vglue -0.4cm
\caption{${\cal O}(N_f\alpha^2)$
         corrections to the photon vacuum polarization,
         $\Pi^{\prime(2)}_{\gamma\gamma}(0)$.
        }
\vglue -0.4cm
\end{figure}

\tracingcommands=1
The result of calculation of these diagrams yields
\begin{align}
\Pi_{\gamma\gamma}^{\prime(2)}&(0)=
\left(\frac{g^2 s_\theta}{16\pi^2}\right)^2 \frac{8}{3}
\frac{(n+2)}{n}(\pi M_W^2)^{n-4}
\Gamma(4-n)\Gamma\left(2-\frac{n}{2}\right)
\Gamma\left(\frac{n}{2}-1\right)
\nonumber\\
&+\left(\frac{g^2 s_\theta}{16\pi^2}\right)^2
\frac{4}{27c_\theta^2}(44 s_\theta^4-27s_\theta^2+9)\nonumber\\
& \ \ \ \ \ \ \ \ \times
\frac{(n-6)}{n}(\pi M_Z^2)^{n-4}
\Gamma(5-n)\Gamma\left(2-\frac{n}{2}\right)
\Gamma\left(\frac{n}{2}-1\right)
\nonumber\\
&-\left(\frac{g^2s_\theta}{16\pi^2}\right)^2
\frac{4(n-2)}{3n}\pi^{n-4}(M_W^2)^{\frac{n}{2}-2}
\Gamma\left(3-\frac{n}{2}\right)\Gamma\left(2-\frac{n}{2}\right)
\nonumber\\& \ \ \ \ \ \ \ \ \times
\sum_f Q_f^2(m_f^2)^{\frac{n}{2}-2}
\nonumber\\
&-\left(\frac{g^2s_\theta}{16\pi^2}\right)^2
\frac{8}{3}\pi^{n-4}(M_W^2)^{\frac{n}{2}-2}
\Gamma\left(2-\frac{n}{2}\right)^2
\sum_f Q_f t_{3f}(m_f^2)^{\frac{n}{2}-2}
\nonumber\\
&-\left(\frac{g^2s_\theta}{16\pi^2}\right)^2
\frac{16(n-2)}{3n}\pi^{n-4}(M_Z^2)^{\frac{n}{2}-2}
\Gamma\left(3-\frac{n}{2}\right)\Gamma\left(2-\frac{n}{2}\right)
\nonumber\\& \ \ \ \ \ \ \ \ \times
\sum_f Q_f^2\left(\frac{\beta_{Lf}^2+\beta_{Rf}^2}{2}\right)
       (m_f^2)^{\frac{n}{2}-2}
\nonumber\\
&+\left(\frac{g^2s_\theta}{16\pi^2}\right)^2
\frac{16n}{3(n-2)}\pi^{n-4}(M_Z^2)^{\frac{n}{2}-2}
\Gamma\left(3-\frac{n}{2}\right)\Gamma\left(2-\frac{n}{2}\right)
\nonumber\\& \ \ \ \ \ \ \ \ \times
\sum_f Q_f^2\beta_{Lf}\beta_{Rf}(m_f^2)^{\frac{n}{2}-2}
\nonumber\\
&+\left(\frac{g^2s_\theta}{16\pi^2}\right)^2
\frac{4s_\theta^2}{3}
\frac{(5n^2-33n+34)}{n(n-5)}\pi^{n-4}
\Gamma\left(3-\frac{n}{2}\right)\Gamma\left(2-\frac{n}{2}\right)
\nonumber\\& \ \ \ \ \ \ \ \ \times
\sum_f Q_f^4 (m_f^2)^{n-4}
\label{eq:twoloopphotonse}
\end{align}
where $\beta_{Lf}$ and $\beta_{Rf}$
are the left- and right-handed couplings of the $Z^0$ to a fermion,
of charge, $Q_f$, weak isospin, $t_{3f}$, and mass, $m_f$, given by
\[
\beta_{Lf}=\frac{t_{3f}-s_\theta^2 Q_f}{c_\theta},\ \ \ \ \ \
\beta_{Rf}=-\frac{s_\theta^2 Q_f}{c_\theta}.
\]

Contributions that are suppressed by factors $m_f^2/M_W^2$ relative to the
leading terms have been dropped.
The first term on the right hand side of eq.(\ref{eq:twoloopphotonse})
comes from diagrams Fig.2a--e that contain an internal $W$ boson, the
second comes from diagrams Fig.2f\&g containing an internal $Z^0$.
The last term in eq.(\ref{eq:twoloopphotonse}) comes from
diagrams Fig.2h\&i and is pure QED
in nature. Contributions for which the fermion mass can be safely
set to zero without affecting the final result were obtained
using the methods
described in ref.\cite{loweng}. The terms in which the fermion mass
appears are obtained using the asymptotic expansion of ref.\cite{DavyTausk}.
Note that setting $m_f=0$ in Fig.2c--g does not
immediately cause any
obvious problems in the computation because the diagram still contains
one non-vanishing scale. A certain amount of care is thus required
to identify situations in which the fermion mass cannot be discarded.

Although the calculation of the photon vacuum polarization is somewhat
lengthy one has the non-trivial check that longitudinal part vanishes.

Dependence on the fermion mass, $m_f$, is eliminated in all diagrams,
with the exception of Fig.2h\&i, by fermion mass
counterterms. The diagrams of Fig.2h\&i
obviously become non-perturbative when the internal
fermions are light quarks. In that case the hadronic contribution is
treated in the usual manner by writing
\[
\Pi_{\gamma\gamma}^{\prime(f)}(0)
=\Re\Pi_{\gamma\gamma}^{\prime(f)}(\hat q^2)
-[\Re\Pi_{\gamma\gamma}^{\prime(f)}(\hat q^2)
    -\Pi_{\gamma\gamma}^{\prime(f)}(0)]
\]
with $\hat q^2$ being chosen to be sufficiently large that perturbative
QCD can be used. The term in square brackets on the rhs can be obtained
in the usual way using dispersion relations and we have calculated for
$|\hat q^2|\gg m_f^2$
\vglue -0.4cm
\begin{multline}
\Pi_{\gamma\gamma}^{\prime(2{\rm QED})}(\hat q^2)=
-\sum_f\left(\frac{g^2}{16\pi^2}\right)^2
Q_f^4 s_\theta^4(\pi\hat q^2)^{n-4}\\
\times
\Bigg\{8(n^2-7n+16)
     \frac{
          \Gamma\left(2-\frac{\displaystyle n}{\displaystyle 2}\right)^2
          \Gamma\left(\frac{\displaystyle n}{\displaystyle 2}\right)^3
          \Gamma\left(\frac{\displaystyle n}{\displaystyle 2}-2\right)
          }
          {\Gamma(n)\Gamma(n-1)}\\
+24\frac{(n^2-4n+8)}{(n-1)(n-4)}.
     \frac{
          \Gamma(4-n)
          \Gamma\left(\frac{\displaystyle n}{\displaystyle 2}\right)^2
          \Gamma\left(\frac{\displaystyle n}{\displaystyle 2}-2\right)
          }
          {\Gamma\left(\frac{\displaystyle 3n}{\displaystyle 2}-2\right)}
\Bigg\}.
\label{eq:highqphotonse}
\end{multline}

Despite appearances, the expression on the right hand side of
eq.(\ref{eq:highqphotonse}) has only a simple pole with a constant
coefficient at $n=4$ that can be canceled by local counterterms.
The leading logarithmic expressions can be found in
ref.\cite[section 8-4-4]{ItzyksonZuber} where the authors invite the
``foolhardy reader'' to check that the finite parts are transverse.
Here we have
gone further and demonstrated this property in the exact result.
Eq.(\ref{eq:highqphotonse}) could also be obtained by applying
analytic continuation
relations for the hypergeometric functions, ${}_2F_1$ and ${}_3F_2$,
to formulas given by Broadhurst {\it et al.}\ \cite{Broadhurst}.

% ---- Bibliography ----
%

\end{document}